\documentclass[twocolumn]{aastex63}

\usepackage{color}

\begin{document}

\title{WISEA J041451.67$-$585456.7 and WISEA J181006.18$-$101000.5: The First Extreme T-type Subdwarfs?}

\correspondingauthor{Adam C. Schneider}
\email{aschneid10@gmail.com}

\author[0000-0002-6294-5937]{Adam C. Schneider}
\affil{School of Earth and Space Exploration, Arizona State University, Tempe, AZ, 85282, USA}

\author[0000-0002-6523-9536]{Adam J. Burgasser}
\affil{Center for Astrophysics and Space Science, University of California San Diego, La Jolla, CA 92093, USA}

\author{Roman Gerasimov}
\affil{Center for Astrophysics and Space Science, University of California San Diego, La Jolla, CA 92093, USA}

\author[0000-0001-7519-1700]{Federico Marocco}
\affil{Jet Propulsion Laboratory, California Institute of Technology, 4800 Oak Grove Dr., Pasadena, CA 91109, USA}
\affil{IPAC, Mail Code 100-22, Caltech, 1200 E. California Blvd., Pasadena, CA 91125, USA}

\author[0000-0002-2592-9612]{Jonathan Gagn\'e}
\affil{Institute for Research on Exoplanets, Universit\'e de Montr\'eal, 2900 Boulevard \'Edouard-Montpetit Montr\'eal, QC H3T~1J4, Canada}
\affil{Plan\'etarium Rio Tinto Alcan, Espace pour la Vie, 4801 ave. Pierre-de Coubertin, Montr\'eal, QC H1V~3V4, Canada}

\author[0000-0003-2236-2320]{Sam Goodman}
\affil{Backyard Worlds: Planet 9, USA}

\author{Paul Beaulieu}
\affil{Backyard Worlds: Planet 9, USA}

\author{William Pendrill}
\affil{Backyard Worlds: Planet 9, USA}

\author[0000-0003-4083-9962]{Austin Rothermich}
\affil{Backyard Worlds: Planet 9, USA}

\author[0000-0003-4864-5484]{Arttu Sainio}
\affil{Backyard Worlds: Planet 9, USA}

\author[0000-0002-2387-5489]{Marc J. Kuchner}
\affil{Exoplanets and Stellar Astrophysics Laboratory, NASA Goddard Space Flight Center, 8800 Greenbelt Road, Greenbelt, MD 20771, USA}

\author[0000-0001-7896-5791]{Dan Caselden}
\affil{Gigamon Applied Threat Research, 619 Western Avenue, Suite 200, Seattle, WA 98104, USA}

\author[0000-0002-1125-7384]{Aaron M. Meisner}
\affil{NSF's National Optical-Infrared Astronomy Research Laboratory, 950 N. Cherry Ave., Tucson, AZ 85719, USA}

\author[0000-0001-6251-0573]{Jacqueline K. Faherty}
\affil{Department of Astrophysics, American Museum of Natural History, Central Park West at 79th Street, NY 10024, USA}

\author[0000-0003-2008-1488]{Eric E. Mamajek}
\affil{Jet Propulsion Laboratory, California Institute of Technology, 4800 Oak Grove drive, Pasadena CA 91109, USA}
\affil{Department of Physics \& Astronomy, University of Rochester, Rochester, NY 14627, USA}

\author[0000-0002-5370-7494]{Chih-Chun Hsu}
\affil{Center for Astrophysics and Space Science, University of California San Diego, La Jolla, CA 92093, USA}

\author[0000-0002-4649-1568]{Jennifer J. Greco}
\affil{Department of Physics and Astronomy, University of Toledo, 2801 West Bancroft St., Toledo, OH 43606, USA}

\author[0000-0001-7780-3352]{Michael C. Cushing}
\affil{Department of Physics and Astronomy, University of Toledo, 2801 West Bancroft St., Toledo, OH 43606, USA}

\author[0000-0003-4269-260X]{J. Davy Kirkpatrick}
\affil{IPAC, Mail Code 100-22, Caltech, 1200 E. California Blvd., Pasadena, CA 91125, USA}

\author[0000-0001-8170-7072]{Daniella Bardalez Gagliuffi}
\affil{Department of Astrophysics, American Museum of Natural History, Central Park West at 79th Street, NY 10024, USA}

\author[0000-0002-9632-9382]{Sarah E. Logsdon}
\affil{NSF's National Optical-Infrared Astronomy Research Laboratory, 950 N. Cherry Ave., Tucson, AZ 85719, USA}

\author[0000-0003-0580-7244]{Katelyn Allers}
\affil{Bucknell University; Department of Physics and Astronomy; Lewisburg, PA 17837, USA}

\author[0000-0002-1783-8817]{John H. Debes}
\affil{ESA for AURA, Space Telescope Science Institute, 3700 San Martin Drive, Baltimore, MD 21218, USA
}

\author{The Backyard Worlds: Planet 9 Collaboration}
\affil{Backyard Worlds: Planet 9, USA}

\begin{abstract}

We present the discoveries of WISEA J041451.67$-$585456.7 and WISEA J181006.18$-$101000.5, two low-temperature (1200--1400 K), high proper motion T-type subdwarfs. Both objects were discovered via their high proper motion ($>$0\farcs5 yr$^{-1}$); WISEA J181006.18$-$101000.5 as part of the NEOWISE proper motion survey and WISEA J041451.67$-$585456.7 as part of the citizen science project Backyard Worlds; Planet 9.   We have confirmed both as brown dwarfs with follow-up near-infrared spectroscopy.  Their spectra and near-infrared colors are unique amongst known brown dwarfs, with some colors consistent with L-type brown dwarfs and other colors resembling those of the latest-type T dwarfs.  While no forward model consistently reproduces the features seen in their near-infrared spectra, the closest matches suggest very low metallicities ([Fe/H] $\leq$ $-$1), making these objects likely the first examples of extreme subdwarfs of the T spectral class (esdT). WISEA J041451.67$-$585456.7 and WISEA J181006.18$-$101000.5 are found to be part of a small population of objects that occupy the ``substellar transition zone,'' and have the lowest masses and effective temperatures of all objects in this group.   

\end{abstract}


\section{Introduction}

\begin{figure*}
\plotone{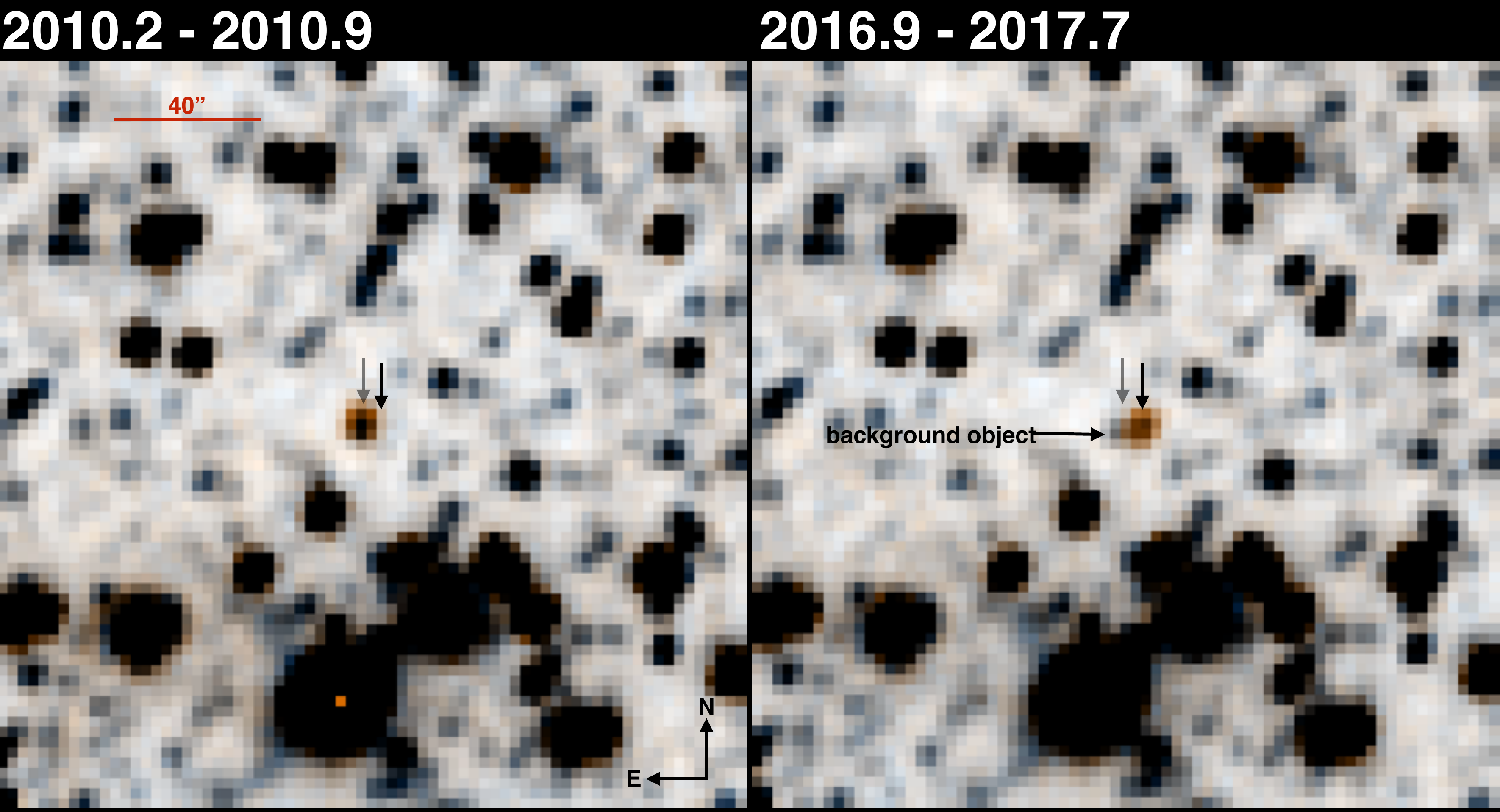}
\caption{unWISE images of WISEA 1810$-$1010 from the WiseView tool (\href{http://byw.tools/wiseview}{byw.tools/wiseview}). On the left is a combination of the first two {\it WISE} epochs pre-hibernation, and on the right is the fifth (NEOWISER) epoch \citep{meis19}. The stationary background source that is blended with WISEA 1810$-$1010 in the first epochs (left panel) can be seen clearly in the latest epochs (right panel).  In this composite image, W1 image is color coded blue and the W2 image is color coded orange.  The orange color of WISEA 1810$-$1010 indicates that the source is brighter at 4.5 $\mu$m (W2) than 3.6 $\mu$m (W1), typical of late-type brown dwarfs. }  
\end{figure*}

Cool subdwarfs are metal-deficient stars and brown dwarfs that make up the Galaxy's earliest generations, and are thus some of the oldest members of the Galactic population.  The low-mass stellar (M-type) subdwarf population has been extensively characterized (e.g., \citealt{gizis97,lepine07}).  There are three recognized subclasses of low-mass M-type subdwarfs: subdwarfs (sd), extreme subdwarfs (esd), and ultra-subdwarfs (usd), which roughly correspond to metallicity values ([Fe/H]) of $-$0.5, $-$1.0, and $-$1.5 dex, respectively \citep{pav15}, though these classes are still undergoing refinements \citep{lod19,zhangs19}.  The intermediate subdwarf class (d/sd; [Fe/H] $>$ $-$0.5) has also been suggested \citep{mould78,burg07}.  Modern large-area infrared surveys, such as the Two Micron All Sky Survey (2MASS; \citealt{skrut06}), have allowed for the extension of subdwarf studies to the L spectral class, beginning with the discovery of the first L-type subdwarf, 2MASS 05325346$+$8246465 \citep{burg03}, originally classified as an sdL7 and more recently reclassified to be part of the extreme subdwarf class (esdL7; \citealt{kirk10}, \citealt{zhang17a}).  As the population of L-type subdwarfs has increased in number, the sdL, esdL, and usdL subclasses have emerged for which \cite{zhang17a} make similar [Fe/H] definitions: sdL are defined as objects with [Fe/H] between $-$0.3 and $-$1.0; esdL have even lower metallicities, between $-$1.0 and $-$1.7; and usdLs are the most metal-poor, with [Fe/H] $\leq$$-$1.7.  

The population of T-type subdwarfs has historically been very small and has been made up of either unusually blue, high tangential velocity objects such as 2MASSI J0937347$+$293142 \citep{burg03,burg06}, ULAS J1233$+$1219 \citep{murray11}, and ULAS J131610.28$+$075553.0 \citep{burn14}, and companions to higher-mass stars for which low metallicities are known \citep{burn10,scholz10,pinfield12,mace13}. Classification of these sources as bona fide subdwarfs has been challenging. For example, the spectra of two early sdT candidates WISE J071121.36$-$573634.2 and WISE J210529.08$-$623558.7, which show thick-disk-like kinematics, possess weak or ambiguous signs of low-metallicity, which may be attributable to multiplicity effects \citep{luhman14, kellogg18}. Even the tangential velocity estimates of these sources, estimated from spectral type--absolute magnitude relations, may be inaccurate for subdwarf types, depending on the band used. \cite{zhang19} have identified a sample of $\geq$T5 objects that could qualify as members of an sdT spectral class, while \cite{greco19} have found three promising early-type sdT candidates. No members of more metal-poor esdT or usdT classes have yet been identified. Extending the metallicity sequence of the T class is essential for a proper assessment of metallicity effects on these atmospheres, which influence cloud formation and evolution, gas chemistry, and the role of pressure-dependent opacity sources such as collision-induced H$_2$. Moreover because of their relatively old age, the vast majority of halo brown dwarfs have likely cooled off to become metal-poor T subdwarfs \citep{burg04}.

Low-metallicity subdwarfs are commonly found through surveys for high proper motion objects. Subdwarfs, which are among the oldest stars in the Galaxy, tend to have kinematics distinct from other stellar populations, either because they have had more time to dynamically interact with Galactic structures that perturb their orbits or because they formed prior to the Galaxy's current disk structure. Both cases result in high velocities relative to disk stars like the Sun \citep{wielen77,gizis97,faherty09}.  Long time baseline images from the {\it Wide-field Infrared Survey Explorer} {\it (WISE)} are now available, allowing for the first all-sky searches for objects with large proper motions at wavelengths between 3 and 5 $\mu$m, where brown dwarf emission peaks (e.g., \citealt{kirk14,kirk16,schneid16}).  Here we report the discovery of the high proper motion brown dwarfs WISEA J041451.67$-$585456.7 and WISEA J181006.18$-$101000.5.  The unusual near-infrared spectra of these sources, and their corresponding temperature and metallicity estimates, suggest that they are likely the first examples of extreme-type subdwarfs of the T spectral class.

\section{Discovery of WISEA J181006.18$-$101000.5 and WISEA J041451.67$-$585456.7}

WISEA J181006.18$-$101000.5 (WISEA 1810$-$1010 hereafter) was initially identified as a potential high proper motion source as part of the NEOWISE proper motion survey \citep{schneid16,greco19}.  However, at that time, its proper motion could not be confirmed because of the high density of background sources near this object ($b$$\approx$4$\fdg$3) and the lack of additional detections at different epochs in other catalogs.  Since the survey of \cite{schneid16}, which had a time baseline of $\sim$1 yr, {\it WISE} has continued to scan the sky as part of the NEOWISE project \citep{mainzer14}, leading to much longer time baselines that can distinguish objects with significant proper motions.  There are also new efforts to identify such sources including the Backyard Worlds: Planet 9 citizen science project (BYW; \citealt{kuch17}) and the CatWISE motion survey \citep{eisen19,mar19}.

The WiseView tool (\href{http://byw.tools/wiseview}{byw.tools/wiseview}; \citealt{cas18}) was created through the BYW project as a resource for visually confirming high proper motion sources using multi-epoch unWISE images \citep{meis18}.  We re-examined WISEA 1810$-$1010 using WiseView and  it was found to have significant motion compared to background sources, confirming it as a high proper motion object.  The images show significant blending with a background source in the pre-hibernation WISE epochs, which becomes much less severe in later post-reactivation (NEOWISER) epochs (Figure 1).  After we examined WISEA 1810$-$1010 with WiseView, we chose it as a high-priority target for follow-up observations because of its large proper motion ($>$1\arcsec yr$^{-1}$) and unusual infrared colors (Table 1).  Additionally, citizen scientist Arttu Sainio independently discovered this object via the BYW project.

WISEA J041451.67$-$585456.7 (WISEA 0414$-$5854 hereafter) was found as part of the BYW project independently by four different citizen scientists: Paul Beaulieu, Sam Goodman, William Pendrill, and Austin Rothermich.  The BYW was launched in 2017 February as part of the Zooniverse citizen science community \citep{kuch17}.  Since then, over 150,000 users around the world have participated in the visual identification and classification of new brown dwarf candidates through the BYW interface\footnote{\url{www.BackyardWorlds.org}}.  The project has discovered the lowest binding energy ultracool binary in the field population \citep{faherty20}, one of the coldest known brown dwarfs (Bardalez Gagliuffi et al.~submitted), and the oldest white dwarf with a dusty disk \citep{debes19}.  WISEA 0414$-$5854 was flagged as a high-priority target for follow-up spectroscopy because it had unusual infrared colors, very similar to those of WISEA 1810$-$1010.

\begin{figure*}
\plotone{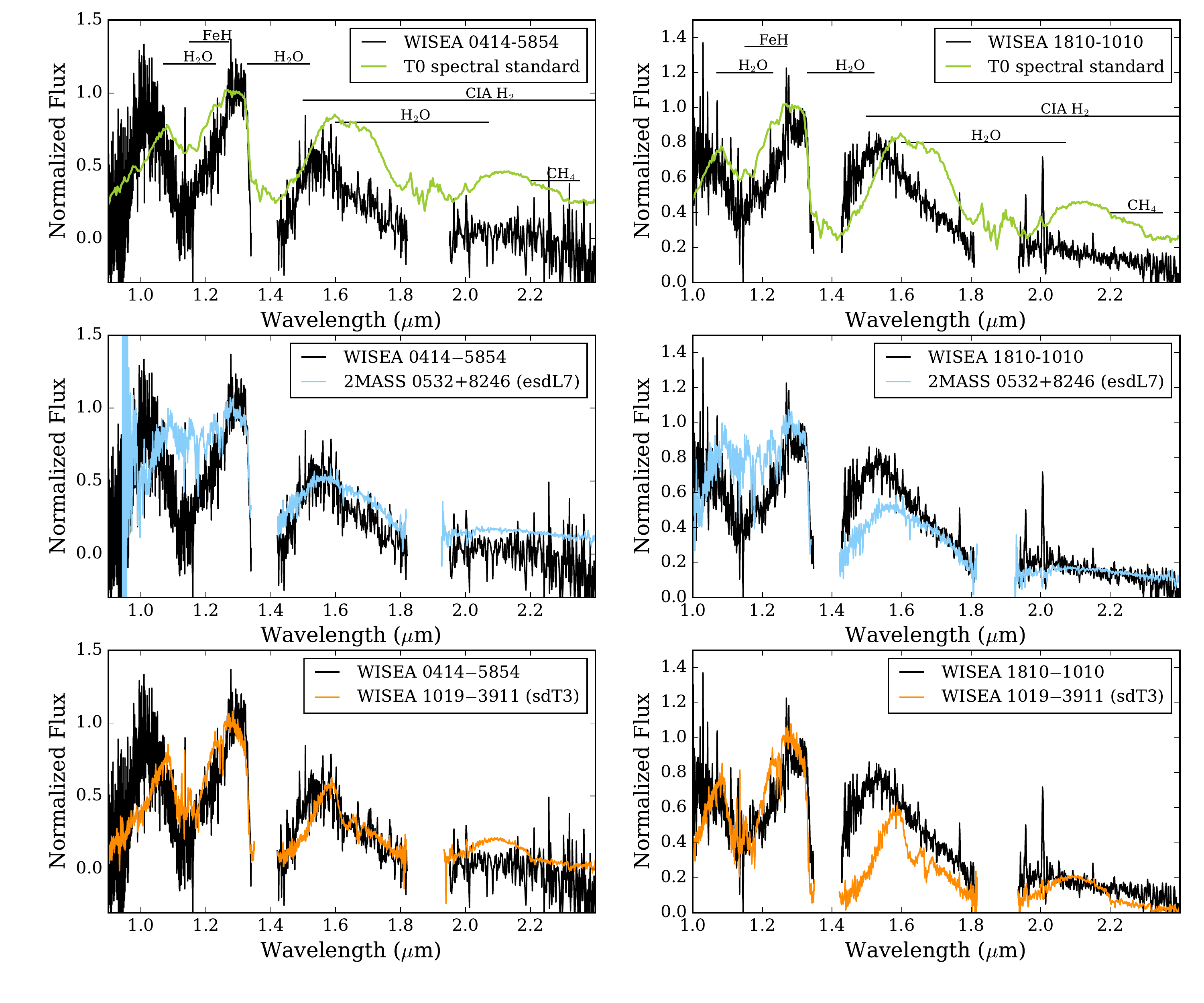}
\caption{The near-infrared spectra of WISEA  0414$-$5854 (black, left panels) and WISEA 1810$-$1010 (black, right panels) compared to the T0 spectral standard (SDSS J120747.17$+$024424.8; \citealt{loop07}), 2MASS 05325346$+$8246465 (esdL7), and WISEA 101944.62$-$391151.6 (sdT3).  All spectra are normalized to the J-band peak between 1.27 and 1.29 $\mu$m.  Major absorbing species are labeled in the upper panels.}  
\end{figure*}

\begin{deluxetable*}{lcccc}
\tablecaption{Properties of WISEA 0414$-$5854 and WISEA 1810$-$1010}
\tablehead{
\colhead{Parameter} & \colhead{Value} & \colhead{Value} & \colhead{Ref.}}
\startdata
\cutinhead{Identifiers}
WISEA & J041451.67$-$585456.7 & J181006.18$-$101000.5 &   1\\
CWISEP & J041451.75$-$585454.0 & J181006.00$-$101001.1 & 2\\
\cutinhead{Observed Properties}
$\mu$$_{\alpha}$ (mas yr$^{-1}$) & 214.0 $\pm$ 29.3 & $-$1113 $\pm$ 12 & 2,3\\
$\mu$$_{\delta}$ (mas yr$^{-1}$)  & 653.8 $\pm$ 26.6 & $-$206 $\pm$ 13 & 2,3\\
$J$ (mag) & 19.632 $\pm$ 0.111 & 17.264 $\pm$ 0.020  & 4,5\\
$H$ (mag) & \dots & 16.500 $\pm$ 0.018  & -,5\\
$K$ (mag) & $>$18.82 & 17.162 $\pm$ 0.081  & 4,5\\
W1 (mag) & 16.705 $\pm$ 0.029 & 13.650 $\pm$ 0.018   & 2,3\\
W2  (mag) & 15.286 $\pm$ 0.025 & 12.483 $\pm$ 0.009  & 2,3\\
\cutinhead{Inferred Properties}
$T_{\rm eff}$ (K)  & 1300 $\pm$ 100 & 1300 $\pm$ 100   & 3\\
log g  & 5.5 $\pm$ 0.5 & 5.0 $\pm$ 0.5 & 3\\
Mass ($M_{\odot}$)  & 0.075--0.080 & 0.075--0.080   & 3\\
$[$Fe/H$]$ (dex) & $\approx$ $-$1 & $\leq$ $-$1 & 3\\
\enddata
\tablerefs{ (1) AllWISE \citep{cutri13}, (2) CatWISE \citep{eisen19}, (3) this work, (4) VHS \citep{mcmahon12}, (5) UKIDSS GPS \citep{law07,lucas08}. }
\end{deluxetable*}

\section{Observations}

\subsection{Keck/NIRES}
We attempted to obtain a spectrum of WISEA 1810$-$1010 with the Near-Infrared Echellette Spectrometer (NIRES; \citealt{wilson04}) on the Keck II telescope on 2019 August 13.  Due to the combination of humid and cloudy weather conditions and WISEA 1810$-$1010's high proper motion in a very crowded field, we were unable to acquire a spectrum of this source at that time.  However, we obtained $K-$band guider images of the field around WISEA 1810$-$1010, which later allowed us to unambiguously pinpoint its current position for subsequent observations with Palomar/TripleSpec.

\subsection{Palomar/TripleSpec}
We observed WISEA 1810$-$1010 with the TripleSpec instrument \citep{herter08} on the Palomar 200 inch telescope on 2019 September 19.  Nine 300 s exposures were taken using the 1\arcsec $\times$ 30\arcsec slit in an ABBA pattern for a total on source time of 2700 s.  The spectrum covers a wavelength range of $\approx$1.0--2.4 $\mu$m at a resolution R $\approx$ 2600.  Because of high humidity, the telescope was closed for approximately 30 minutes after the last on-target exposure.  An A0-type telluric standard was observed immediately after the telescope was reopened.  Data were reduced using a modified version of the SpeXTool package \citep{vacca03,cush04}.  The final reduced spectrum is shown in the left panels of Figure 2.

\subsection{Magellan/FIRE}
We observed WISEA 0414$-$5854 with the Folded-port Infrared Echellette (FIRE; \citealt{simcoe13}) spectrograph at the 6.5 m Baade Magellan telescope on 2020 February 12.  Observations were taken using the 0\farcs6 slit in prism mode which achieves a resolving power between 500 and 300 across the $J$, $H$, and $K$ bands (0.8--2.5 $\mu$m).  We used the sample-up-the-ramp mode, obtaining 12 exposures of 126.8 s, for a total on source exposure time of $\sim$1500 s.  An A0V star was observed immediately after for telluric correction purposes.  Reductions were performed with a modified version of the SpeXTool package \citep{vacca03,cush04}, and the final reduced spectrum is shown in the right panels of Figure 2.

\section{Analysis}

\begin{figure*}
\plotone{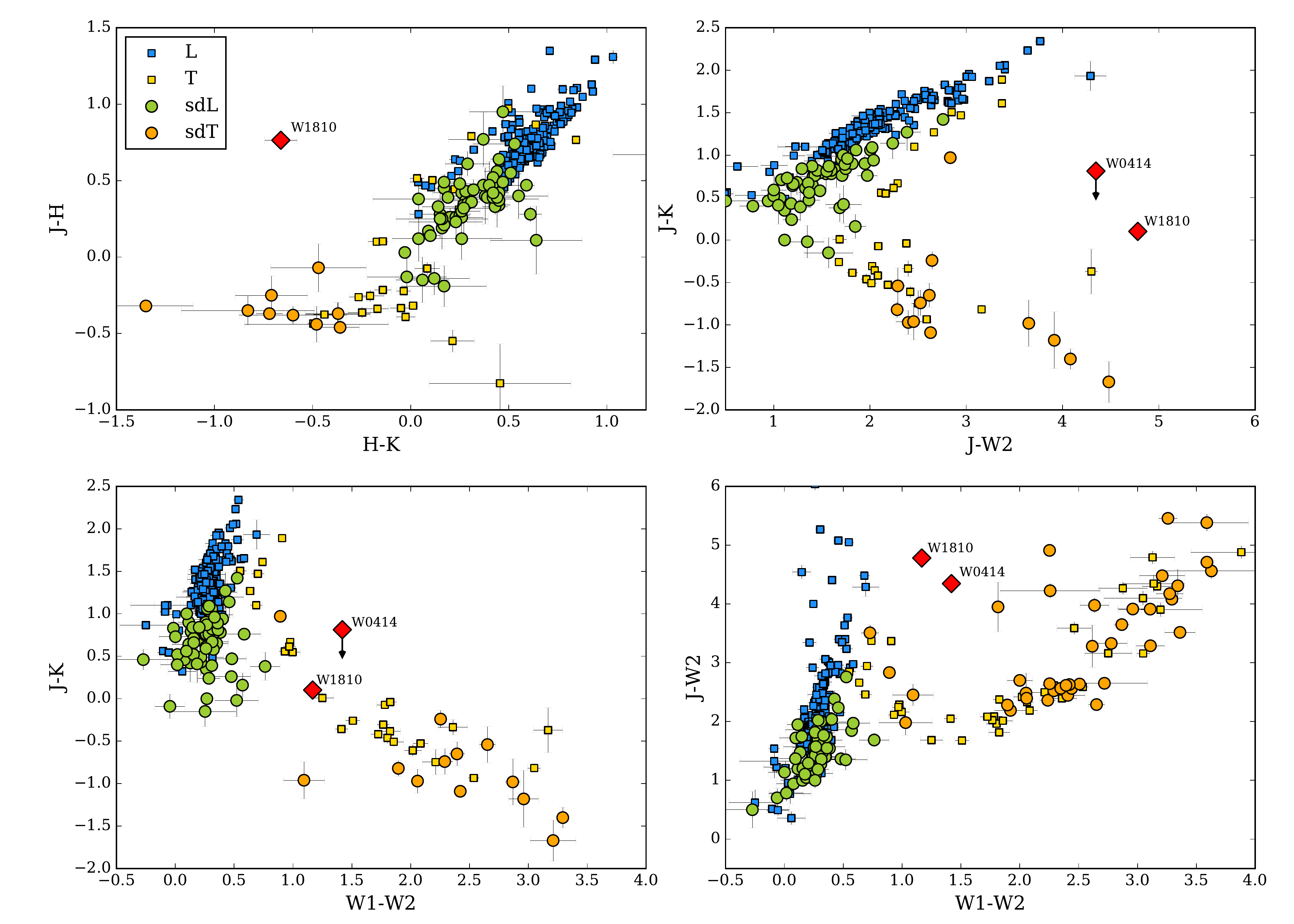}
\caption{Color--color diagrams comparing WISEA 0414$-$5854 and WISEA 1810$-$1010 (red diamonds) to known field and subdwarf L and T dwarfs.  Field L and T dwarfs primarily come from \cite{burn13}, \cite{day13}, and \cite{mar15} while known L and T subdwarfs come from \cite{zhang18b} and  \cite{zhang19}.  Marker definitions are provided in the legend in the first panel.  WISEA 0414$-$5854 is not plotted in the first panel because no $H-$band magnitude is currently available for this object.}  
\end{figure*}

\subsection{Photometry}

WISEA 0414$-$5854 was detected by the Vista Hemisphere Survey (VHS; \citealt{mcmahon12}), with a secure detection at $J$, but a nondetection at $K$ (and not observed at $H$). We determined a $K-$band limit by querying the VHS database with a large radius (300\arcsec) around the position of WISEA 0414$-$5854 and fitting a fourth-order polynomial to the measured magnitudes and corresponding signal-to-noise ratios (S/N).  We find a 3$\sigma$ $K-$band limit of 18.82 mag.  WISEA 0414$-$5854 also has a clear, unblended detection in the CatWISE preliminary catalog.  Photometry for WISEA 0414$-$5854 is given in Table 1.

WISEA 1810$-$1010 was detected by the UKIRT Infrared Deep Sky Survey (UKIDSS; \citealt{law07}) as part of the Galactic Plane Survey (GPS; \citealt{lucas08}).  $JHK$ photometry from the UKIDSS GPS is listed in Table 1.  The only other survey to detect WISEA 1810$-$1010 was {\it WISE},  however, as noted previously and seen in Figure 1, WISEA 1810$-$1010 is blended with a background object in the earliest {\it WISE} epochs.  For this reason, we do not use photometry from the {\it WISE} All-Sky, AllWISE, or CatWISE catalogs, as each of these uses the initial {\it WISE} epochs to measure photometry.  Instead, we apply the CatWISE pipeline \citep{eisen19} to only the NEOWISE epochs of this source.  The resulting photometry is listed in Table 1.  For reference, the {\it WISE} All-Sky, AllWISE, and CatWISE catalog photometry values are on average $\sim$0.19 mag brighter in W1, with only marginal differences between measured W2 magnitudes ($<$2$\sigma$).

Figure 3 shows various color-color diagrams comparing the infrared colors of WISEA 0414$-$5854 and WISEA 1810$-$1010 to known field L and T dwarfs detected in UKIDSS along with the compilations of known sdLs in \cite{zhang18b} and proposed sdTs in \cite{zhang19}.  As seen in the figure, the colors of these two objects are unique among brown dwarfs of all types.  The upper left panel shows that WISEA 1810$-$1010 has a $J-H$ color consistent with the L dwarf population, but a blue $H-K$ color that is consistent with the sdT population.  The W1-W2 colors of of WISEA 0414$-$5854 and WISEA 1810$-$1010 would typically correspond to spectral types of $\sim$T3--T4, however, their $J-$W2 colors are consistent with either the latest-type L dwarfs or the latest-type T dwarfs as seen in the bottom right panel of the figure.  Note that \cite{meisner20} used {\it Spitzer Space Telescope} photometry and found several objects with anomalously red $J-$ch2 colors compared to their ch1$-$ch2 colors, where the ch1 and ch2 {\it Spitzer} passbands are very similar to W1 and W2 from {\it WISE}.  These objects with similar colors as WISEA 0414$-$5854 and WISEA 1810$-$1010 may be additional members of this population.  

\subsection{Astrometry}
As mentioned previously, WISEA 1810$-$1010 is blended with a background source in the initial {\it WISE} epochs.  To determine the proper motion of WISEA 1810$-$1010, we ran the CatWISE pipeline \citep{eisen19} omitting the first two epochs where the astrometry is likely affected by the unmoving background source.  We find $\mu_{\alpha}$ = -1333$\pm$12 mas yr$^{-1}$ and $\mu_{\delta}$ = -206$\pm$13 mas yr$^{-1}$.  Note that the proper motion values in the CatWISE preliminary catalog are $\mu_{\alpha}$ = -901.6$\pm$10.3 mas yr$^{-1}$ and $\mu_{\delta}$ = -155$\pm$10.5 mas yr$^{-1}$.  Users of the preliminary CatWISE catalog should keep in mind this potential issue in astrometry for sources in heavily crowded fields \citep{eisen19}.   The proper motion of  WISEA 0414$-$5854 was taken from the CatWISE preliminary catalog and is listed in Table~1.

\subsection{Spectral Types/Model Fitting}
L and T spectral types in the infrared are typically determined by comparing the full 1--2.5 $\mu$m spectrum or individual portions to sets of spectral standards \citep{burg06,kirk10,cruz18}.  The T0 spectral standard (SDSS J120747.17$+$024424.8; \citealt{loop07}) is the best match to both WISEA 0414$-$5854 and WISEA 1810$-$1010 in the $J-$band; however the $H$ and $K$ fluxes are depressed relative to peak in $J$, as compared with the standard (Figure 2, top panels). This behavior is similar to what is seen in the latest L-type subdwarfs, where the spectra are relatively featureless and the infrared flux peaks blueward of the peak location typically seen for field L dwarfs at $H-$band \citep{gonz18}. This shift is most likely due to collision-induced absorption (CIA) from H$_2$, which broadly absorbs across the near-infrared with a peak in absorption at the fundamental vibration frequency at 4161 cm$^{-1}$ (2.4 $\mu$m; \citealt{linsky69}; \citealt{bory97}; \citealt{burg03}). There is a also a significant discrepancy between our observed spectra and the T0 standard spectrum around the 1.15 $\mu$m H$_2$O band, which is considerably broader in the former. This may be related to previously identified peculiarities in the shapes of 1.05 $\mu$m spectral peak observed in ``mild'' T subdwarfs such as 2MASS J0937$+$2931 and ULAS J141623.94$+$134836.3 \citep{burg06,burg10}. 

\begin{figure*}
\plotone{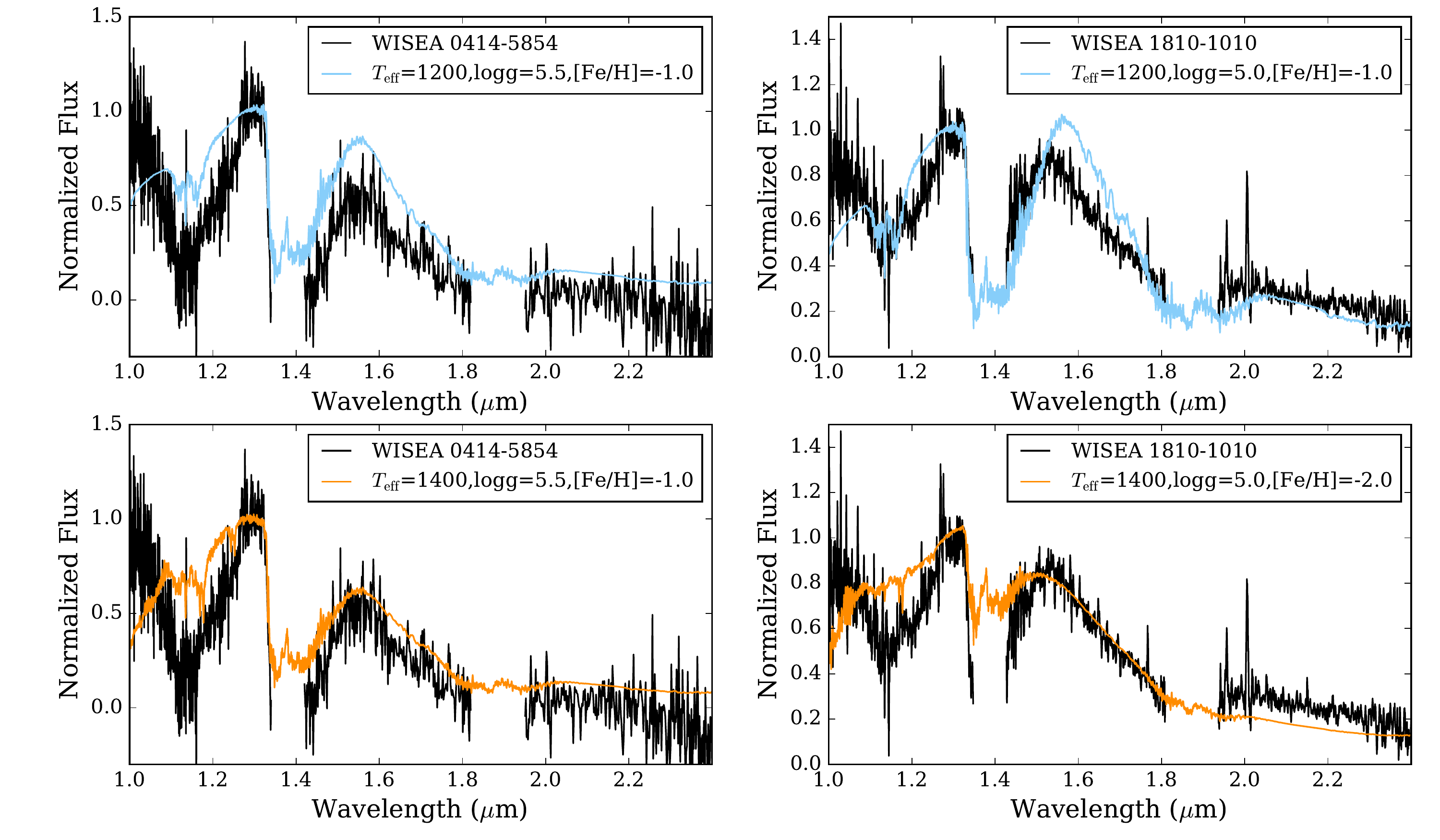}
\caption{Near-infrared spectra of WISEA  0414$-$5854 (black, left panels) and WISEA 1810$-$1010 (black, right panels) compared to two low-temperature PHOENIX models.  All spectra are normalized to the J-band peak between 1.27 and 1.29 $\mu$m.}  
\end{figure*}

The comparison of the spectra of WISEA 0414$-$5854 and WISEA 1810$-$1010 to spectral standards thus suggests these may be early-T type subdwarfs.  As mentioned in Section 1, early-T type subdwarfs are rare and often show weak or ambiguous signs of low-metallicity, making comparisons to known early-T type subdwarfs challenging.  To verify this, we compare in Figure~2 the spectra of WISEA 0414$-$5854 and WISEA 1810$-$1010 to that of the known esdL7 2MASS 05325346$+$8246465 \citep{burg03, kirk10, zhang17a} and to that of the suspected sdT3 WISEA 101944.62$-$391151.6 \citep{greco19}.   When compared to 2MASS 05325346$+$8246465, both WISEA 0414$-$5854 and WISEA 1810$-$1010 have much deeper and broader absorption around 1.15 $\mu$m.  The $H-$band shapes of both objects, however, show similarities to the esdL7 source by being relatively featureless, though WISEA 0414$-$5854 and WISEA 1810$-$1010 have a steeper slope on the red side of the $H-$band peak.  The $J-$band spectral shapes of WISEA 0414$-$5854 and WISEA 1810$-$1010 look more similar to that of the sdT3 object, especially the depth of the absorption feature at 1.15$\mu$m.  However, we note that WISEA 0414$-$5854 has a peak flux in the $Y-$band $\sim$0.1 $\mu$m bluer than that of WISEA 101944.62$-$391151.6.  The spectrum of WISEA 1810$-$1010 does not extend blueward of 1 $\mu$m, but shows a $Y-$band morphology similar to that of WISEA 0414$-$5854.  These comparisons are strongly suggestive of early-T subdwarf or extreme subdwarf spectral classifications. 

To estimate the physical properties of WISEA 0414$-$5854 and WISEA 1810$-$1010 ($T_{\rm eff}$, log g, [Fe/H]), models were generated using a subbranch of the PHOENIX v15 code (\citealt{haus99,allard13}, Gerasimov et al.\ in prep) which includes dust condensation and gravitational settling.  A grid was produced for $T_{\rm eff}$ values from 900 to 1500 K in steps of 100 K, [Fe/H] values from 0 to $-$3 dex in steps of 1 dex, and log g values of 5.0 and 5.5.  All models are publicly available\footnote{http://atmos.ucsd.edu/?p=atlas}. No model is able to accurately reproduce all of WISEA 0414$-$5854's or WISEA 1810$-$1010's spectral features.  The two best fitting models for each were found using a simple $\chi$$^2$ statistic and are shown in Figure~4.  

For WISEA 0414$-$5854, the best matching models imply a $T_{\rm eff}$ of 1200--1400 K, a log(g) of 5.5, and an [Fe/H] value of $-$1.  However, there are significant differences between the observed spectrum and the best-fitting models, particularly in the $J-$band wavelength range.  For WISEA1810$-$1010, the $T_{\rm eff}$ = 1200 K model reproduces some features of the $J$-band portion of its spectrum and is a reasonable match at $K$, but substantially overpredicts the $H$-band flux.  Alternatively, the $T_{\rm eff}$ = 1400 K model does not match as well at $J$ and underpredicts the $K$-band flux, but matches very well in the $H$-band portion of the spectrum.  Based on the best-fitting models, we estimate a $T_{\rm eff}$ of 1300$\pm$100 K and [Fe/H] $\leq$ $-$1 dex for WISEA 0414$-$5854 and WISEA 1810$-$1010.  

The spectral morphologies of these sources, along with [Fe/H] estimates $\leq$ $-$1 dex, suggest extreme subdwarf spectral subclass designations are appropriate for both of these objects.  We note that while the best-fitting models for WISEA 0414$-$5854 have [Fe/H] values of $-$1, which is the nominal boundary between the sd and esd classes, the similarities of the spectral morphologies of WISEA 0414$-$5854 and WISEA 1810$-$1010 suggest they should have the same subclass.  $T_{\rm eff}$ estimates of 1200--1400 K correspond to spectral types of L7 to T1.5 for field type brown dwarfs \citep{fil15}, in agreement with the best matching spectral standard.  Note, however, that the effective temperatures of late-type sdMs and sdLs are generally warmer than their field counterparts for a given spectral subtype (e.g., \citealt{zhang17a,gonz18}), although it is unclear if this trend extends into the T dwarf regime.  We therefore tentatively assign a spectral type of esdT0$\pm$1 for both of these objects.    

\section{Discussion}

\subsection{Kinematics}

Subdwarfs are typically found to have large space motions relative to the Sun, and belong either to the thick disk or halo populations.  To further investigate the kinematics of WISEA 0414$-$5854 and WISEA 1810$-$1010, two additional measurements are needed: radial velocities and distances.  The spectra presented here are too low in S/N, and lack sharp features, to obtain robust radial velocities. Such measures may be obtained with future high-resolution spectroscopic observations on a larger aperture facility.  Likewise, no parallax is currently available for either object, each object being too faint to be detected in {\it Gaia}.  A distance estimate would allow for a measurement of the tangential velocities ($V_{\rm tan}$) of WISEA 0414$-$5854 and WISEA 1810$-$1010, which could be used to help decide whether they belong to the thin disk, thick disk, or halo populations (e.g., \citealt{bensby03,dup12}).  

Photometric distances are often used when direct distance measurements are not available.  This is problematic for both of these unusual objects.  First, photometric distances typically require a spectral type, and the spectral types of WISEA 0414$-$5854 and WISEA 1810$-$1010 are highly uncertain (see the previous section).  Second, photometric distances are typically constructed for the field population of brown dwarfs and therefore may not be appropriate for subdwarfs.  Those relations that do pertain to subdwarfs either do not yet extend to the T dwarf regime \citep{schil09,zhang17a, gonz18} or are only appropriate for sdT type brown dwarfs with spectral types later than T5 \citep{zhang19}.  If we proceed with absolute magnitude--spectral type relations for field brown dwarfs \citep{dup12} and designate the spectral type of WISEA 1810$-$1010 to be T0, wildly discrepant distances are found for different passbands.  We find a distance of $\sim$14 pc using WISEA 1810$-$1010's W2 magnitude, but a distance of $\sim$67 pc using the $K-$band magnitude.  Combined with our measured proper motion (Table 1), these distance estimates lead to a  $V_{\rm tan}$ range of 77--360 km s$^{-1}$.  For WISEA 0414$-$5854, a similarly large distance range is found (52--94 pc), corresponding to a $V_{\rm tan}$ range of 170--307 km s$^{-1}$.  Parallax measurements for these objects will be necessary to determine accurate $V_{\rm tan}$ values.  Nevertheless, it is clear that both sources are high velocity objects consistent with membership in the metal-poor Galactic thick disk or halo.

\subsection{The Space Density of Early-type esdTs}
The discoveries of WISEA 0414$-$5854 and WISEA 1810$-$1010 allow us to place a crude lower limit on the space density of early-type esdTs.  This limit compared to the space density of early T dwarfs in the field that are not metal-poor can provide a preliminary sense for how common extremely metal-poor objects are in this temperature regime. Two significant sources of uncertainty on the esdT limit are the spectral type and distance estimates for these two new objects.  However, if we take our best spectral type estimates at face value and conservatively choose the further edge of our distance estimates ($\sim$90 pc for WISEA 0414$-$5854), we derive a lower limit for the space density of early-type esdTs of 6.55$\pm$4.63 $\times$ 10$^{-7}$ pc$^{-3}$.  The number of single T0$-$T2 (inclusive) brown dwarfs within 20 pc is 25 (Kirkpatrick et al.~in prep), leading to a space density of 7.46$\pm$1.49 $\times$ 10$^{-3}$ pc$^{-3}$.  This implies that early-type esdTs make up (at least) 0.08\% of the early T dwarf population.  Alternatively, if we take the distance of WISEA 0414$-$5854 to be 50 pc, the esdT fraction would be at least 0.51\%.

There have been many estimates of the space density of thick disk ($f_{\rm TD}$) and halo ($f_{\rm h}$) stars in the solar neighborhood.  Early estimates for $f_{\rm h}$ based on star counts have ranged from 0.06 to 0.2\% (e.g., \citealt{gould98,robin03}), though more recent estimates based on spectroscopic studies and kinematic data from the {\it Gaia} mission have converged on values $\sim$0.45\% (e.g., \citealt{posti18,amarante20}).   Similarly, the values of $f_{\rm TD}$ have varied, with the most recent estimates ranging from 4 to 7\% (e.g., \citealt{bland16,amarante20}).  Considering WISEA 0414$-$5854 and WISEA 1810$-$1010 likely belong to the thick disk or halo population, if we take 0.45\% (7\%) as the true fraction of early T dwarfs that belong to the esdT class, the expected number of esdTs within 90 pc is $\sim$10 ($\sim$159).  If the detection limit for esdTs is closer to 50 pc, the predicted number of esdTs would be $\sim$2 ($\sim$27) for an early-type esdT fraction of 0.45\% (7\%). It is therefore possible that numerous additional early-type esdTs are waiting to be found in the solar neighborhood.  These values assume that the stellar $f_{\rm TD}$ and $f_h$ values hold for substellar members of the solar neighborhood.  A dedicated search for more of these objects could resolve whether or not the stellar values of $f_{\rm TD}$ and $f_h$ are applicable in this temperature regime.  
        
\subsection{WISEA 0414$-$5854, WISEA 1810$-$1010, and the Substellar Transition Zone}
As previously noted in the literature, the higher rates of cooling of metal-poor brown dwarfs \citep{bar97} and the divergent evolution of stars and brown dwarfs over time can give rise to a temperature-dependent ``gap'' across a metallicity-dependent hydrogen-burning mass limit \citep{burg04}. Such a gap has already been detected in globular clusters with the terminus of the hydrogen-burning main sequence \citep{bedin01} and the first candidate globular cluster brown dwarfs \citep{dieball16}. Among nearby low-temperature subdwarfs, evidence of a ``subdwarf gap'' has also started to emerge \citep{kirk14,zhang17b}. In fact, this ``gap'' is more appropriately described as an underdensity of sources across a stellar--substellar transition phase, as hydrogen fusion does not abruptly cease at the hydrogen-burning limit. Sources with reduced fusion can eventually cool to achieve thermal stability, albeit after billions of years (see \citealt{burg08} for the case of 2MASS 05325346$+$8246465). \cite{zhang17b} have explicitly defined a ``substellar transition zone,'' covering a narrow range of masses and large range of effective temperatures over which ``unstable fusion'' can occur. The masses encompassed by this transition, and the range of temperatures it spans depend on the metallicity and age of the population, with a more pronounced spread expected for metal-poor halo subdwarfs. \cite{zhang18a} have designated nine objects as residing in this transition zone, all of which belong to either the esdL or usdL spectral classes.

Figure 5 shows effective temperature and metallicity values for all known L and T subdwarfs with published $T_{\rm eff}$ and [Fe/H] measurements. Along with all of the ``transition zone'' L dwarfs, WISEA 0414$-$5854 and WISEA 1810$-$1010 are on the substellar side of the stellar/substellar boundary at their metallicities.  Based on the metallicity-dependent evolutionary models of \cite{burrows98}, WISEA 0414$-$5854 and WISEA 1810$-$1010 have predicted masses between 0.075 and 0.08 $M_{\odot}$, giving them the lowest masses and temperature estimates of all transitional subdwarfs. 

\begin{figure*}
\plotone{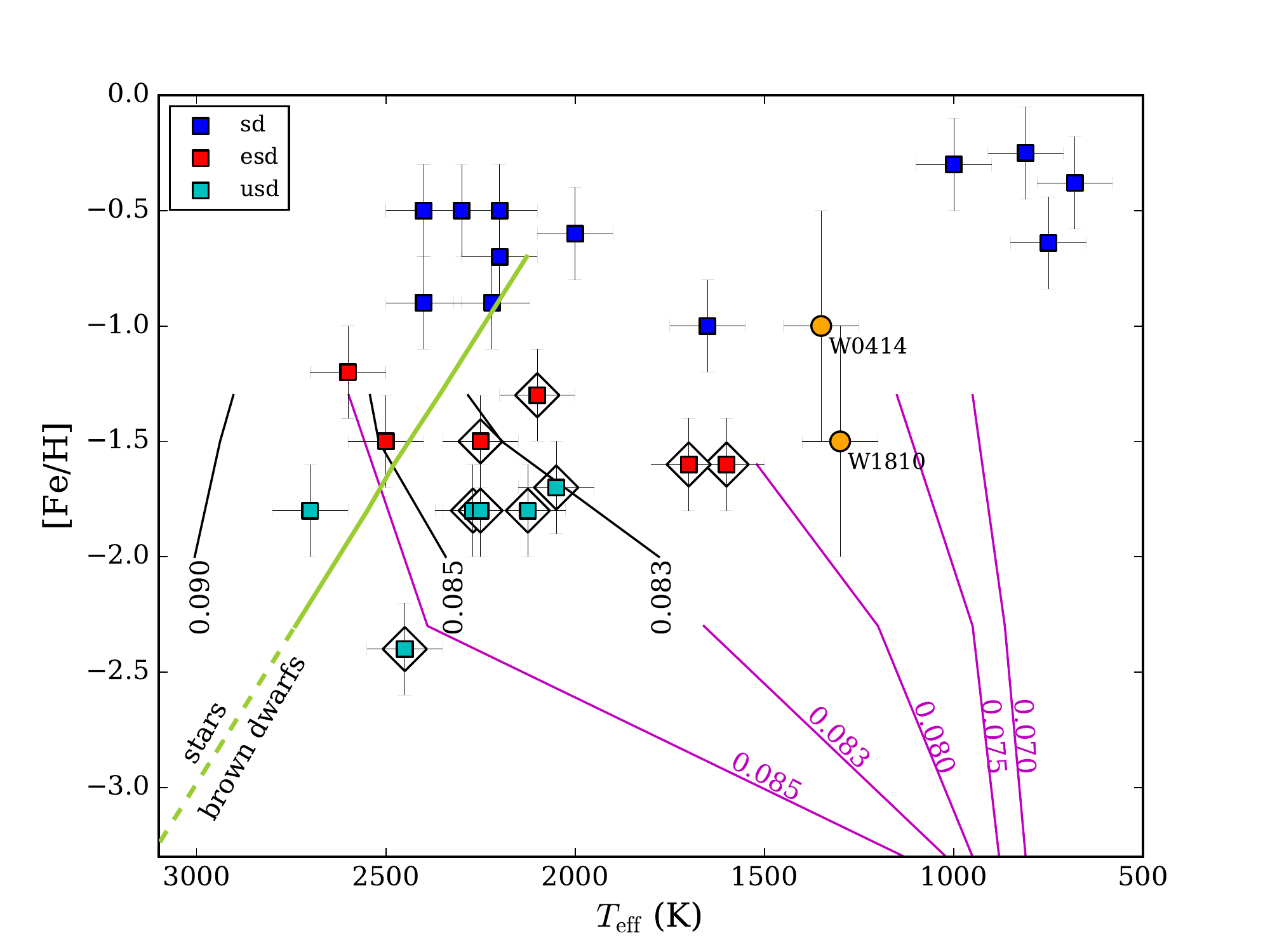}
\caption{[Fe/H] versus $T_{\rm eff}$ for L and T subdwarfs compared to WISEA 0414$-$5854 and WISEA 1810$-$1010 (adapted from \cite{zhang17b}).  Black lines trace 10 Gyr isomass models from \cite{bar97} with masses labeled, while magenta lines trace isomass models from \cite{burrows98}.  Metallicities and effective temperatures for known L subdwarfs come from \cite{zhang17a,zhang17b,zhang18a} and T subdwarfs are from \cite{burg06}, \cite{murray11}, \cite{pinfield12}, and \cite{mace13}.  Objects in the ``substellar transition zone'' from \cite{zhang18a} are highlighted with black diamonds and the hydrogen-burning minimum mass from \cite{zhang17b} is shown as a green line. }
\end{figure*}

\section{Conclusion}

We find that the high proper motion objects WISEA 0414$-$5854 and WISEA 1810$-$1010 have exceptionally unusual spectroscopic and photometric properties that likely reflect significantly subsolar metallicities.  The best-fitting models for these objects suggest very low metallicities ([Fe/H] $\leq$ $-$1), though no single model provides a satisfactory fit across all wavelengths.  Further astrometric and spectroscopic observations are warranted to better characterize these enigmatic systems.  

These discoveries highlight one of the key advantages of using proper motion instead of colors to identify cold, low-luminosity objects.  Color searches are often constructed to find ultracool dwarfs with solar-like metallicities, and are thereby biased against detecting objects with unusual colors such as WISEA 0414$-$5854 and WISEA 1810$-$1010.

\acknowledgments
This research was supported by NASA Astrophysics Data Analysis Program grant NNH17AE75I.  The Backyard Worlds: Planet 9 team would like to thank the many Zooniverse volunteers who have participated in this project, from providing feedback during the beta review stage to classifying flipbooks to contributing to the discussions on TALK.  This work used the Extreme Science and Engineering Discovery Environment (XSEDE) Comet cluster at the San Diego Supercomputer Center (program AST190045). XSEDE is supported by National Science Foundation grant number ACI-1548562 \citep{towns14}.  This publication makes use of data products from the {\it Wide-field Infrared Survey Explorer}, which is a joint project of the University of California, Los Angeles, and the Jet Propulsion Laboratory/California Institute of Technology, and NEOWISE which is a project of the Jet Propulsion Laboratory/California Institute of Technology. {\it WISE} and NEOWISE are funded by the National Aeronautics and Space Administration.  Part of this research was carried out at the Jet Propulsion Laboratory, California Institute of Technology, under a contract with the National Aeronautics and Space Administration.


\begin{thebibliography}{}

\bibitem[Allard et al.(2013)]{allard13} Allard, F., Homeier, D., Freytag, B., et al.\ 2013, Memorie della Societa Astronomica Italiana Supplementi, 24, 128
\bibitem[Amarante et al.(2020)]{amarante20} Amarante, J.~A.~S., Smith, M.~C., \& Boeche, C.\ 2020, \mnras, 492, 3816
\bibitem[Debes et al.(2019)]{debes19} Debes, J.~H., Th{\'e}venot, M., Kuchner, M.~J., et al.\ 2019, \apjl, 872, L25
\bibitem[Baraffe et al.(1997)]{bar97} Baraffe, I., Chabrier, G., Allard, F., et al.\ 1997, \aap, 327, 1054
\bibitem[Bedin et al.(2001)]{bedin01} Bedin, L.~R., Anderson, J., King, I.~R., et al.\ 2001, \apjl, 560, L75
\bibitem[Bensby et al.(2003)]{bensby03} Bensby, T., Feltzing, S., \& Lundstr{\"o}m, I.\ 2003, \aap, 410, 527
\bibitem[Bland-Hawthorn \& Gerhard(2016)]{bland16} Bland-Hawthorn, J., \& Gerhard, O.\ 2016, \araa, 54, 529
\bibitem[Borysow et al.(1997)]{bory97} Borysow, A., Jorgensen, U.~G., \& Zheng, C.\ 1997, \aap, 324, 185
\bibitem[Burgasser et al.(2003)]{burg03} Burgasser, A.~J., Kirkpatrick, J.~D., Burrows, A., et al.\ 2003, \apj, 592, 1186
\bibitem[Burgasser(2004)]{burg04} Burgasser, A.~J.\ 2004, \apjs, 155, 191
\bibitem[Burgasser et al.(2006)]{burg06} Burgasser, A.~J., Burrows, A., \& Kirkpatrick, J.~D.\ 2006, \apj, 639, 1095
\bibitem[Burgasser et al.(2007)]{burg07} Burgasser, A.~J., Cruz, K.~L., \& Kirkpatrick, J.~D.\ 2007, \apj, 657, 494
\bibitem[Burgasser et al.(2008)]{burg08} Burgasser, A.~J., Vrba, F.~J., L{\'e}pine, S., et al.\ 2008, \apj, 672, 1159
\bibitem[Burgasser et al.(2010)]{burg10} Burgasser, A.~J., Looper, D., \& Rayner, J.~T.\ 2010, \aj, 139, 2448
\bibitem[Burningham et al.(2010)]{burn10} Burningham, B., Leggett, S.~K., Lucas, P.~W., et al.\ 2010, \mnras, 404, 1952
\bibitem[Burningham et al.(2013)]{burn13} Burningham, B., Cardoso, C.~V., Smith, L., et al.\ 2013, \mnras, 433, 457
\bibitem[Burningham et al.(2014)]{burn14} Burningham, B., Smith, L., Cardoso, C.~V., et al.\ 2014, \mnras, 440, 359
\bibitem[Burrows et al.(1998)]{burrows98} Burrows, A., Sudarsky, D., Sharp, C., et al.\ 1998, Brown Dwarfs and Extrasolar Planets, 354
\bibitem[Burrows et al.(2001)]{burrows01} Burrows, A., Hubbard, W.~B., Lunine, J.~I., et al.\ 2001, Reviews of Modern Physics, 73, 719
\bibitem[Caselden et al.(2018)]{cas18} Caselden, D., Westin, P., Meisner, A., et al.\ 2018, WiseView: Visualizing motion and variability of faint WISE sources, ascl:1806.004
\bibitem[Cutri et al.(2013)]{cutri13} Cutri, R.~M., et al.\ 2013, yCat, 2328, 0
\bibitem[Cruz et al.(2018)]{cruz18} Cruz, K.~L., N{\'u}{\~n}ez, A., Burgasser, A.~J., et al.\ 2018, \aj, 155, 34
\bibitem[Cushing et al.(2004)]{cush04} Cushing, M.~C., Vacca, W.~D., \& Rayner, J.~T.\ 2004, \pasp, 116, 362
\bibitem[Day-Jones et al.(2013)]{day13} Day-Jones, A.~C., Marocco, F., Pinfield, D.~J., et al.\ 2013, \mnras, 430, 1171
\bibitem[Dieball et al.(2016)]{dieball16} Dieball, A., Bedin, L.~R., Knigge, C., et al.\ 2016, \apj, 817, 48
\bibitem[Dupuy, \& Liu(2012)]{dup12} Dupuy, T.~J., \& Liu, M.~C.\ 2012, \apjs, 201, 19
\bibitem[Eisenhardt et al.(2019)]{eisen19} Eisenhardt, P.~R.~M., Marocco, F., Fowler, J.~W., et al.\ 2019, arXiv e-prints, arXiv:1908.08902
\bibitem[Faherty et al.(2009)]{faherty09} Faherty, J.~K., Burgasser, A.~J., Cruz, K.~L., et al.\ 2009, \aj, 137, 1
\bibitem[Faherty et al.(2020)]{faherty20} Faherty, J.~K., Goodman, S., Caselden, D., et al.\ 2020, \apj, 889, 176
\bibitem[Filippazzo et al.(2015)]{fil15} Filippazzo, J.~C., Rice, E.~L., Faherty, J., et al.\ 2015, \apj, 810, 158
\bibitem[Gizis(1997)]{gizis97} Gizis, J.~E.\ 1997, \aj, 113, 806
\bibitem[Gonzales et al.(2018)]{gonz18} Gonzales, E.~C., Faherty, J.~K., Gagn{\'e}, J., et al.\ 2018, \apj, 864, 100
\bibitem[Gould et al.(1998)]{gould98} Gould, A., Flynn, C., \& Bahcall, J.~N.\ 1998, \apj, 503, 798
\bibitem[Greco et al.(2019)]{greco19} Greco, J.~J., Schneider, A.~C., Cushing, M.~C., et al.\ 2019, \aj, 158, 182
\bibitem[Hauschildt \& Baron(1999)]{haus99} Hauschildt, P.~H., \& Baron, E.\ 1999, Journal of Computational and Applied Mathematics, 109, 41
\bibitem[Herter et al.(2008)]{herter08} Herter, T.~L., Henderson, C.~P., Wilson, J.~C., et al.\ 2008, \procspie, 70140X
\bibitem[Kellogg et al.(2018)]{kellogg18} Kellogg, K., Kirkpatrick, J.~D., Metchev, S., et al.\ 2018, \aj, 155, 87
\bibitem[Kirkpatrick et al.(2010)]{kirk10} Kirkpatrick, J.~D., Looper, D.~L., Burgasser, A.~J., et al.\ 2010, \apjs, 190, 100
\bibitem[Kirkpatrick et al.(2014)]{kirk14} Kirkpatrick, J.~D., Schneider, A., Fajardo-Acosta, S., et al.\ 2014, \apj, 783, 122
\bibitem[Kirkpatrick et al.(2016)]{kirk16} Kirkpatrick, J.~D., Kellogg, K., Schneider, A.~C., et al.\ 2016, \apjs, 224, 36
\bibitem[Kuchner et al.(2017)]{kuch17} Kuchner, M.~J., Faherty, J.~K., Schneider, A.~C., et al.\ 2017, \apjl, 841, L19
\bibitem[Lawrence et al.(2007)]{law07} Lawrence, A., Warren, S.~J., Almaini, O., et al.\ 2007, \mnras, 379, 1599
\bibitem[L{\'e}pine et al.(2007)]{lepine07} L{\'e}pine, S., Rich, R.~M., \& Shara, M.~M.\ 2007, \apj, 669, 1235
\bibitem[Linsky(1969)]{linsky69} Linsky, J.~L.\ 1969, \apj, 156, 989
\bibitem[Lodieu et al.(2019)]{lod19} Lodieu, N., Allard, F., Rodrigo, C., et al.\ 2019, \aap, 628, A61
\bibitem[Looper et al.(2007)]{loop07} Looper, D.~L., Kirkpatrick, J.~D., \& Burgasser, A.~J.\ 2007, \aj, 134, 1162
\bibitem[Lucas et al.(2008)]{lucas08} Lucas, P.~W., Hoare, M.~G., Longmore, A., et al.\ 2008, \mnras, 391, 136
\bibitem[Luhman, \& Sheppard(2014)]{luhman14} Luhman, K.~L., \& Sheppard, S.~S.\ 2014, \apj, 787, 126
\bibitem[Mace et al.(2013)]{mace13} Mace, G.~N., Kirkpatrick, J.~D., Cushing, M.~C., et al.\ 2013, \apj, 777, 36
\bibitem[Mainzer et al.(2014)]{mainzer14} Mainzer, A., Bauer, J., Cutri, R.~M., et al.\ 2014, \apj, 792, 30
\bibitem[Marocco et al.(2015)]{mar15} Marocco, F., Jones, H.~R.~A., Day-Jones, A.~C., et al.\ 2015, \mnras, 449, 3651
\bibitem[Marocco et al.(2019)]{mar19} Marocco, F., Caselden, D., Meisner, A.~M., et al.\ 2019, \apj, 881, 17
\bibitem[McMahon(2012)]{mcmahon12} McMahon, R.\ 2012, Science from the Next Generation Imaging and Spectroscopic Surveys, 37
\bibitem[Meisner et al.(2018)]{meis18} Meisner, A.~M., Lang, D., \& Schlegel, D.~J.\ 2018, \aj, 156, 69
\bibitem[Meisner et al.(2019)]{meis19} Meisner, A.~M., Lang, D., Schlafly, E.~F., et al.\ 2019, \pasp, 131, 124504
\bibitem[Meisner et al.(2020)]{meisner20} Meisner, A.~M., Caselden, D., Kirkpatrick, J.~D., et al.\ 2020, \apj, 889, 74
\bibitem[Mould \& McElroy(1978)]{mould78} Mould, J.~R., \& McElroy, D.~B.\ 1978, \apj, 221, 580
\bibitem[Murray et al.(2011)]{murray11} Murray, D.~N., Burningham, B., Jones, H.~R.~A., et al.\ 2011, \mnras, 414, 575
\bibitem[Pavlenko et al.(2015)]{pav15} Pavlenko, Y.~V., Zhang, Z.~H., G{\'a}lvez-Ortiz, M.~C., et al.\ 2015, \aap, 582, A92
\bibitem[Pinfield et al.(2012)]{pinfield12} Pinfield, D.~J., Burningham, B., Lodieu, N., et al.\ 2012, \mnras, 422, 1922
\bibitem[Posti et al.(2018)]{posti18} Posti, L., Helmi, A., Veljanoski, J., et al.\ 2018, \aap, 615, A70
\bibitem[Robin et al.(2003)]{robin03} Robin, A.~C., Reyl{\'e}, C., Derri{\`e}re, S., et al.\ 2003, \aap, 409, 523
\bibitem[Schilbach et al.(2009)]{schil09} Schilbach, E., R{\"o}ser, S., \& Scholz, R.-D.\ 2009, \aap, 493, L27
\bibitem[Schneider et al.(2016)]{schneid16} Schneider, A.~C., Greco, J., Cushing, M.~C., et al.\ 2016, \apj, 817, 112
\bibitem[Scholz(2010)]{scholz10} Scholz, R.-D.\ 2010, \aap, 515, A92
\bibitem[Simcoe et al.(2013)]{simcoe13} Simcoe, R.~A., Burgasser, A.~J., Schechter, P.~L., et al.\ 2013, \pasp, 125, 270
\bibitem[Skrutskie et al.(2006)]{skrut06} Skrutskie, M.~F., Cutri, R.~M., Stiening, R., et al.\ 2006, \aj, 131, 1163
\bibitem[Towns et al.(2014)]{towns14} Towns, J., Cockerill, T., Dahan, M., et al.\ 2014, CSE, 5, 62
\bibitem[Vacca et al.(2003)]{vacca03} Vacca, W.~D., Cushing, M.~C., \& Rayner, J.~T.\ 2003, \pasp, 115, 389
\bibitem[Wielen(1977)]{wielen77} Wielen, R.\ 1977, \aap, 60, 263
\bibitem[Wilson et al.(2004)]{wilson04} Wilson, J.~C., Henderson, C.~P., Herter, T.~L., et al.\ 2004, \procspie, 1295
\bibitem[Zhang et al.(2019)]{zhangs19} Zhang, S., Luo, A.-L., Comte, G., et al.\ 2019, \apjs, 240, 31
\bibitem[Zhang et al.(2017a)]{zhang17a} Zhang, Z.~H., Pinfield, D.~J., G{\'a}lvez-Ortiz, M.~C., et al.\ 2017a, \mnras, 464, 3040
\bibitem[Zhang et al.(2017b)]{zhang17b} Zhang, Z.~H., Homeier, D., Pinfield, D.~J., et al.\ 2017b, \mnras, 468, 261
\bibitem[Zhang et al.(2018a)]{zhang18a} Zhang, Z.~H., Pinfield, D.~J., G{\'a}lvez-Ortiz, M.~C., et al.\ 2018a, \mnras, 479, 1383
\bibitem[Zhang et al.(2018b)]{zhang18b} Zhang, Z.~H., Galvez-Ortiz, M.~C., Pinfield, D.~J., et al.\ 2018b, \mnras, 480, 5447
\bibitem[Zhang et al.(2019)]{zhang19} Zhang, Z.~H., Burgasser, A.~J., G{\'a}lvez-Ortiz, M.~C., et al.\ 2019, \mnras, 486, 1260

\end{thebibliography}
\end{document}